\documentclass[nonacm,sigconf]{acmart}
\usepackage[utf8]{inputenc}
\usepackage{xspace}
\usepackage{graphicx}
\usepackage{fancyvrb}
\fvset{listparameters=\setlength{\topsep}{5pt}}
\DefineVerbatimEnvironment{mverbatim}{Verbatim}{xleftmargin=10mm}
\usepackage{listings}
\renewcommand{\paragraph}[1]{\vspace*{.5mm} \noindent \textbf{#1}}
\newcommand{\duetsgx}{$\text{\textsc{DuetSGX}}$\xspace}
\newcommand{\duet}{\textsc{Duet}\xspace}

\title{\duetsgx: Differential Privacy with Secure Hardware}

\author{Phillip Nguyen}
\email{Phillip.Nguyen@uvm.edu}
\affiliation{%
  \institution{University of Vermont}
}

\author{Alex Silence}
\email{Alex.Silence@uvm.edu}
\affiliation{%
  \institution{University of Vermont}
}

\author{David Darais}
\email{David.Darais@uvm.edu}
\affiliation{%
  \institution{University of Vermont}
}

\author{Joseph P. Near}
\email{jnear@uvm.edu}
\affiliation{%
  \institution{University of Vermont}
}

\begin{document}

\begin{abstract}
Differential privacy offers a formal privacy guarantee for individuals, but many deployments of differentially private systems require a trusted third party (the data curator). We propose \duetsgx, a system that uses \emph{secure hardware} (Intel's SGX) to eliminate the need for a trusted data curator. Data owners submit encrypted data that can be decrypted \emph{only} within a secure enclave running the \duetsgx system, ensuring that sensitive data is never available to the data curator. Analysts submit queries written in the \duet language, which is specifically designed for verifying that programs satisfy differential privacy; \duetsgx uses the \duet typechecker to verify that each query satisfies differential privacy before running it. \duetsgx therefore provides the benefits of local differential privacy \emph{and} central differential privacy simultaneously: noise is only added to final results, and there is no trusted third party. We have implemented a proof-of-concept implementation of \duetsgx and we release it as open-source.
\end{abstract}

\maketitle

\section{Introduction}

Differential privacy~\cite{dwork2006calibrating, dwork2014algorithmic} and its variants represent the ``gold standard'' in privacy protection, and it has been adopted by organizations like Google~\cite{erlingsson2014rappor}, Apple~\cite{apple}, and the US Census Bureau~\cite{abowd2018us} to protect the privacy of individuals. However, challenges in deploying differential privacy remain.

Key among these challenges is the problem of the \emph{trusted data curator}. In the \emph{central model} of differential privacy~\cite{dwork2014algorithmic}, data owners contribute their sensitive data to a central repository maintained by a \emph{data curator}, who runs differentially private computations on the data and reveals the results. In the central model, the data curator must be \emph{trusted} to safeguard the sensitive data and reveal \emph{only} differentially private results.

What if the data curator is \emph{not trustworthy}, or becomes compromised? In that case, the adversary can simply \emph{look at the data}, and differential privacy provides no benefit at all. To address this problem, the \emph{local model} of differential privacy~\cite{dwork2014algorithmic} requires data contributors to add noise to their data \emph{before} submitting it to the data curator. Since this data already satisfies differential privacy when it is collected, the data curator does not need to be trusted. This approach is used in systems at Apple~\cite{apple} and Google~\cite{erlingsson2014rappor}, due to the advantages of its threat model. Unfortunately, the local model of differential privacy requires \emph{significantly} more noise than the central model, making the results much less accurate.

We study an important question in differential privacy research: \emph{can we achieve the benefits of both models simultaneously?} In other words, our goal is to provide the more accurate answers that are possible in the central model, while eliminating the requirement for a trusted data curator, as in the local model.

This paper presents \duetsgx, a proof-of-concept system that uses \emph{secure hardware} to achieve the accuracy of the central model without the requirement of a trusted data curator. \duetsgx is a platform for collecting sensitive data and performing differentially private queries on that data. The data submitted to a \duetsgx server is encrypted, and computations on the data are protected by a \emph{secure enclave} provided by Intel's Software Guard Extensions (SGX)~\cite{costan2016intel}. The secure enclave prevents other processes---including the operating system---from examining the sensitive data or intermediate results of computation. The enclave also enables \emph{remote attestation}, a protocol by which data owners can verify that they are communicating with a valid \duetsgx server running on a real SGX processor. Data owners must trust the hardware itself (and thus its manufacturer, Intel), but do \emph{not} need to trust the data curator in control of that hardware.

\duetsgx requires that all queries satisfy differential privacy. Raw data can never be extracted by any party from a \duetsgx server. To ensure that queries satisfy differential privacy, \duetsgx requires queries to be specified in the \duet language~\cite{near2019duet}, and uses the \duet typechecker to verify that each query satisfies differential privacy and to determine its privacy cost. If the query satisfies differential privacy and its privacy cost does not exceed the remaining privacy budget, \duetsgx uses the \duet interpreter to execute the query on the encrypted sensitive data. To ensure confidentiality and integrity, \duetsgx runs the \duet typechecker and interpreter inside of the secure enclave.

We have built a proof-of-concept implementation of \duetsgx that provides a RESTful API accessible by standard HTTP requests. Our implementation re-uses the existing implementation of the \duet typechecker and interpreter, and adds capabilities for handling encrypted data and managing a global privacy budget. To run inside of a secure enclave, \duetsgx relies on Graphene~\cite{tsai2017graphene}, which provides the ability to run unmodified programs inside of enclaves. Our proof-of-concept has some important limitations (discussed in Section~\ref{sec:limitations}), but it demonstrates the feasibility of the approach and represents the first step towards a practical, deployable platform. Our implementation is available as open source.\footnote{\url{https://github.com/uvm-plaid/duet-sgx}}

\section{Background}

\paragraph{Differential Privacy.}
Differential privacy is a formal privacy definition that bounds the effect any single individual can have on the outcome of an analysis. Formally, we say that a \emph{mechanism} $\mathcal{M}$ satisfies $(\epsilon, \delta)$-differential privacy if for all databases $x, y \in \mathcal{D}$ with a \emph{distance metric} $d(x,y) \in \mathbb{R}$, and for all possible sets of outcomes $S$, $\text{Pr}[\mathcal{M}(x) \subseteq S] \leq e^\epsilon \text{Pr}[\mathcal{M}(y) \subseteq S] + \delta$ when $d(x,y) <= 1$. We can achieve differential privacy using several basic mechanisms; for example, the \emph{Laplace mechanism} adds noise drawn from the Laplace distribution and satisfies $(\epsilon, 0)$-differential privacy (or ``pure'' differential privacy), and the \emph{Gaussian mechanism} adds Gaussian noise and satisfies $(\epsilon, \delta)$-differential privacy (or ``approximate'' differential privacy). Differential privacy mechanisms are \emph{compositional}: running $\mathcal{M}$ twice on the same data satisfies $(2\epsilon, 2\delta)$-differential privacy. For a detailed summary of differential privacy, see Dwork \& Roth~\cite{dwork2014algorithmic}.

\paragraph{\duet.}
\duet~\cite{near2019duet} is a programming language for writing differentially private programs and verifying correct use of differential privacy. In \duet, privacy is encoded through \emph{types}, and when Duet accepts a program as well-typed, this amounts to a proof that the program will satisfy differential privacy when executed. Duet's typechecker and interpreter were previously implemented as part of the \duet project.\footnote{\url{https://github.com/uvm-plaid/duet}} See Section~\ref{sec:spec-duetsgx-quer} for more information on \duet and how it is used in \duetsgx.

\paragraph{Secure Hardware.}
Intel's Software Guard Extensions (SGX)~\cite{costan2016intel} is a security feature present on most of Intel's recent CPUs. SGX is one example of a \emph{Trusted Execution Environment (TEE)}; other examples include ARM TrustZone and AMD Secure Encrypted Virtualization. Each one provides a slightly different programming interface, but all are intended to provide \emph{secure execution}.

SGX allows programs to launch \emph{secure enclaves}, which protect the confidentiality and integrity of computation occurring within the enclave---even from privileged software like the operating system. To protect against untrusted parties ``faking'' the presence of SGX, it also provides remote attestation capabilities. Remote attestation allows the CPU to produce a \emph{quote} certifying that (1) the software is running inside of an enclave on a legitimate SGX CPU, and (2) the software has not been tampered with. The CPU constructs the quote by computing a digest of the contents of the enclave and signing it using a private key built into the CPU itself. Intel provides an attestation service~\cite{intel_attestation} for verifying the authenticity of these quotes.

% ## Background
%  - Differential privacy
%  - Duet
%  - SGX

\section{\duetsgx Overview}

This section provides an overview of the \duetsgx system: its threat model and goals (Section~\ref{sec:threat-model}), its architecture and API (Section~\ref{sec:architecture}), and its implementation (Section~\ref{sec:implementation}).

\subsection{Threat Model \& Guarantees}
\label{sec:threat-model}

The goal of \duetsgx is to ensure confidentiality for the sensitive data submitted by data owners, and to guarantee that all results computed and output by the system satisfy differential privacy. We consider a setting in which a malicious adversary may control the server running the \duetsgx platform, the network connecting that server to data owners, and the analyst who submits queries to the platform. We assume that a data owner's device cannot be corrupted by the adversary, and that the SGX hardware protections are also robust against the adversary (more on this assumption in Section~\ref{sec:limitations}).

Under these assumptions, \duetsgx ensures confidentiality for submitted data, and differential privacy for all outputs. The submitted data is encrypted so that it can only be processed within a secure enclave, protecting it from the data curator. Queries submitted by the analyst are verified to satisfy differential privacy by the \duet typechecker, preventing a malicious analyst from writing a query to reveal the submitted data.

\subsection{Architecture}
\label{sec:architecture}

Usage of \duetsgx proceeds in two phases: (1) data collection, and (2) query processing.

\begin{figure}
  \centering
  \includegraphics[width=.48\textwidth]{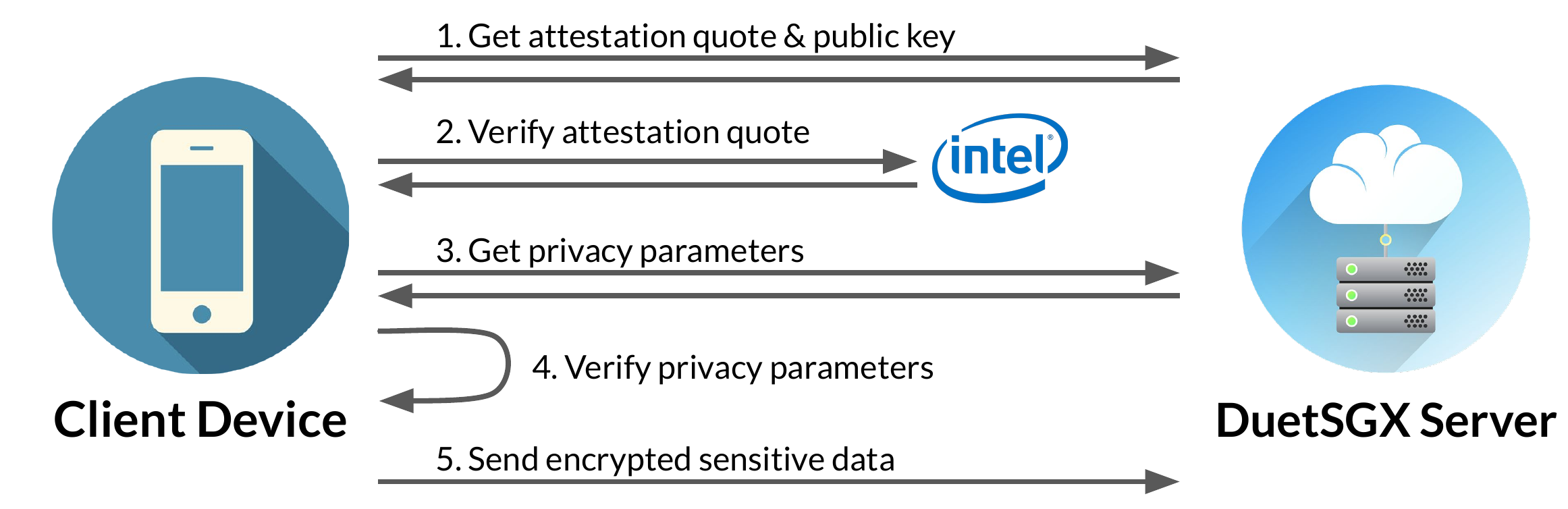}
  \caption{Data Collection Phase. In this phase, the client device negotiates with the \duetsgx server to securely submit data for differentially private analysis.}
  \label{fig:phase_1}
\end{figure}

\paragraph{Phase 1: Data Collection.}
In the first phase, client-side software communicates with the \duetsgx server to contribute data. The \duetsgx server provides a RESTful HTTP interface for submitting data and queries, summarized in Figure~\ref{fig:api}. In general, a data owner submits a data element to the \duetsgx server following the process shown in Figure~\ref{fig:phase_1}. The first step is to obtain an SGX attestation quote from the \duetsgx server and verify it with Intel's attestation service~\cite{intel_attestation}. Second, the client obtains the current privacy budget settings, signed using the \duetsgx server's public key, and verifies that they satisfy the data owner's wishes. Finally, the client encrypts the sensitive data using the server's public key and submits it to the \duetsgx server.

The API is implemented using a Python Flask server and can be reached using standard HTTP requests. The \duetsgx server also provides an HTML interface using Javascript to interact with the API, allowing clients to submit data via a web browser. The web-based interface encrypts data in the browser, using a Javascript implementation of RSA.

The key pair used in this process is generated within the secure enclave when the \duetsgx server starts up, and the private key never leaves the enclave. When the server shuts down, the private key is lost, and the encrypted data that has been submitted so far is effectively destroyed.

\begin{figure}
  \centering
  \includegraphics[width=.48\textwidth]{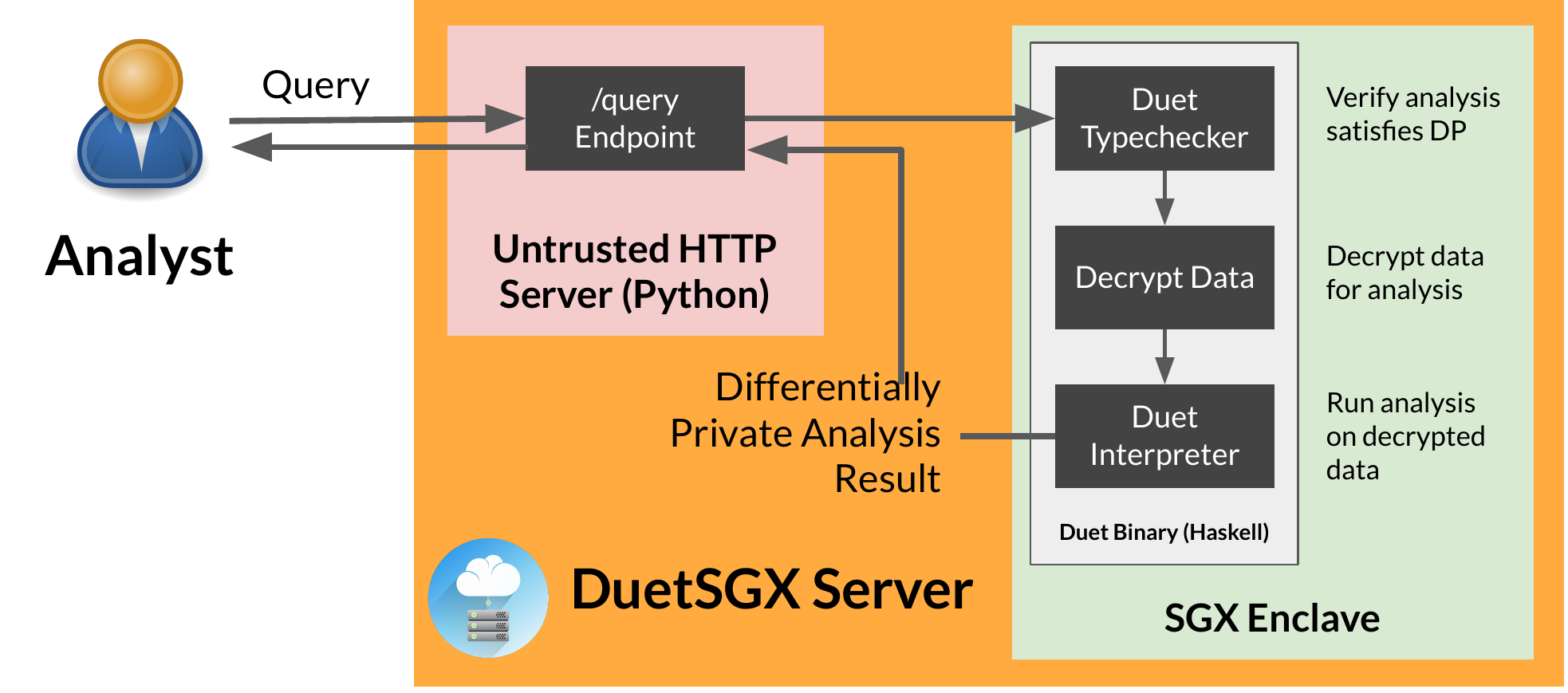}
  \caption{Data Analysis Phase. In this phase, the \duetsgx server runs \duet programs on the submitted data to compute differentially private results.}
  \label{fig:phase_2}
\end{figure}

\paragraph{Phase 2: Query Processing.}
The second phase is processing differentially private queries over the submitted data. Analysts write their queries using the \duet language (detailed in Section~\ref{sec:spec-duetsgx-quer}) and submit them to the \texttt{/query} endpoint of the \duetsgx server. The server runs the query and returns a differentially private result. This process is summarized in Figure~\ref{fig:phase_2}.

The HTTP server itself is a Python program that is \emph{untrusted}. The trusted \duetsgx binary runs entirely inside of an SGX enclave, and interacts with the untrusted HTTP server by reading and writing files on disk. As described earlier, the key pair used for encrypting and decrypting the sensitive data is generated when the enclave starts up, and the private key is stored only in enclave memory; when the \duetsgx server shuts down, the private key is (intentionally) lost.

The privacy budget is set by the data curator when the \duetsgx server starts up by specifying the initial values for $\epsilon$ and $\delta$ in a file. When the server starts, it signs these values using the generated key pair, so that the data curator cannot modify the privacy budget after the server is running. When a query is submitted, the \duet typechecker derives a privacy cost for it and subtracts that amount from the remaining budget. If the budget would be exhausted by the query, then the query is rejected; otherwise, the query's cost is subtracted from the budget and the new value of the remaining budget is signed by the \duetsgx server.

\begin{figure}
\begin{tabular}{l c r}
\textbf{API} & \textbf{Method} & \textbf{Description}\\ \hline
/epsilon & GET &
Returns the current $\epsilon$ value
\\ \hline
/delta & GET &
Returns the current $\delta$ value
\\ \hline
/attest & GET &
Returns the attestation quote
\\ \hline
/pubkeypem & GET &
Returns the public key pem
\\ \hline
/insert & POST &
Insert data into the database
\\ \hline
/query & POST &
Submit a query for execution
\end{tabular}
\caption{\duetsgx Server API}
\label{fig:api}
\end{figure}

% \begin{figure*}
% \centering
% \includegraphics[width=\textwidth]{figures/fig3.PNG}
% \caption{Web interface}
% \label{fig:interface}
% \end{figure*}

\subsection{Implementation}
\label{sec:implementation}

The \duetsgx binary that runs inside of the secure enclave is written in Haskell, and builds on the \duet implementation (which is also written in Haskell). The \duetsgx enclave code implements the specific interface with the untrusted HTTP server, handles the generation of encryption keys and encryption and decryption of data, and coordinates the use of the \duet typechecker to determine privacy cost and the \duet interpreter to compute results. The untrusted HTTP server is a simple Python program implemented using the Flask library.

Intel SGX enclaves are not normally capable of running Haskell programs, so we use the Graphene system~\cite{tsai2017graphene} to enable \duetsgx to run on SGX. Graphene provides a compatibility layer that allows arbitrary Linux programs to run in an SGX enclave. Graphene simplifies the development of systems like \duetsgx, but also brings limitations, especially for remote attestation. We discuss these in Section~\ref{sec:limitations}.

\section{Specifying \duetsgx Queries}
\label{sec:spec-duetsgx-quer}

\duetsgx runs queries specified in \duet~\cite{near2019duet}, a programming language and type system designed for automatically proving that a program satisfies differential privacy. When a query is submitted, the \duetsgx server runs the \duet typechecker \emph{before} running the query, to (1) verify that the submitted query satisfies differential privacy, and (2) determine the privacy cost $(\epsilon, \delta)$ of running the query.

\duet's typechecker is entirely static and does not require access to the data. Its results can therefore safely be used to determine the privacy cost of a query before running it, and to reject queries that do not satisfy differential privacy or have privacy costs greater than the remaining privacy budget.

\paragraph{Writing \duet Programs.}
\duet includes basic primitives for differentially private programming like the Laplace mechanism (written \texttt{laplace}) and the Gaussian mechanism (written \texttt{gauss}). For example, the following program adds Gaussian noise to the variable \texttt{x}, assuming that \texttt{x} has a sensitivity of 1.0:

\begin{lstlisting}[xleftmargin=10mm]
let $\epsilon$ = $\mathbb{R}^{+}$[1.5] in
let $\delta$ = $\mathbb{R}^{+}$[0.000001] in
gauss[$\mathbb{R}^{+}$[1.0], $\epsilon$, $\delta$] <x> { x }
\end{lstlisting}

\noindent The notation \texttt{$\mathbb{R}^{+}$[1.5]} is used to specify a \emph{statically-known} non-negative real value, which the typechecker can use to help determine privacy cost. Sensitivities and the values of privacy parameters must be statically known in order for the \duet typechecker to bound privacy cost, and statically-known values are assumed to be publicly known. The list of variables inside angle brackets (\texttt{<x>}) denote the variables for which the Gaussian mechanism should attempt to provide privacy; variables not listed here will be treated as auxiliary information and assigned infinite privacy cost. The expression inside curly braces (\texttt{\{ x \}}) represents the actual result to which noise will be added.

\duet also allows specifying \emph{privacy functions} (with \texttt{p$\lambda$}) whose types encode their privacy costs. For example:

\begin{lstlisting}[xleftmargin=10mm]
p$\lambda$ . x : $\mathbb{R}$ $\Rightarrow$
  gauss[$\mathbb{R}^{+}$[1.0], $\mathbb{R}^{+}$[1.0], $\mathbb{R}^{+}$[0.001]] <x> { x }
\end{lstlisting}

\noindent The \duet typechecker reports that this program has the type:

\begin{lstlisting}[xleftmargin=10mm]
$\mathbb{R}$@$\langle$1.0, 0.001$\rangle$ $\Rightarrow$ $\mathbb{R}$
\end{lstlisting}

\noindent This type indicates that the program's value is a function which satisfies $(1.0, 0.001)$-differential privacy (the privacy cost is listed after the \texttt{@} sign). For more information on the \duet language, see the \duet paper~\cite{near2019duet} or the project webpage~\cite{duet_page}.

\paragraph{\duet Programs as \duetsgx Queries.}
A \duetsgx query is simply a \duet program with a specific form. A valid \duetsgx query is a program that defines a privacy function of a single argument (the data stored in the secure database). The privacy function must have constant privacy cost and the type of the function's argument must match the schema of the secure database. For example, the following program represents a \duetsgx query that counts the number of rows in a secure database of pairs of real numbers (e.g. latitude/longitude pairs):

\begin{lstlisting}[xleftmargin=10mm]
p$\lambda$ . df :  M [L1 , U | $\bigstar$ , d $\mathbb{R}$ :: d $\mathbb{R}$ :: [] ] $\Rightarrow$
  let $\epsilon$ = $\mathbb{R}^{+}$[1.0] in
  let $\delta$ = $\mathbb{R}^{+}$[0.001] in
  gauss[$\mathbb{R}^{+}$[1.0], $\epsilon$, $\delta$] <df> { real (rows df) }
\end{lstlisting}

\noindent The notation \texttt{M [L$infty$,U|...]} is a \duet \emph{matrix type}, which we use in \duetsgx queries to specify the schema of the database. The \texttt{L1} annotation specifies the distance metric used to define neighboring inputs (we add up the number of differing rows), and \texttt{U} specifies that there is no known upper bound on values in the matrix. \texttt{$\bigstar$} means the number of rows in the matrix is not statically known (this will usually be the case for \duetsgx databases). \texttt{d $\mathbb{R}$ :: d $\mathbb{R}$ :: []} specifies the schema of the matrix, in linked-list notation; this example has two columns, each containing real numbers. \texttt{d $\mathbb{R}$} is the \emph{discrete real number} type, which is the same as a real number except with a boolean distance metric: two values $v_1, v_2 \in$ \texttt{d $\mathbb{R}$} are 0 apart if they are equal, and 1 apart otherwise. \texttt{rows} is a built-in function that counts the number of rows in a matrix, and \texttt{real} converts a discrete real number into a real number. We have used an unrealistically large value for $\delta$ in this example for readability. This program has the type:

\begin{lstlisting}[xleftmargin=10mm]
M [L1,U | $\bigstar$, d $\mathbb{R}$::d $\mathbb{R}$::[]]@$\langle$1.0, 0.001$\rangle$ $\Rightarrow$ $\mathbb{R}$
\end{lstlisting}

\noindent The \duet typechecker returns this type to the \duetsgx implementation, which uses the privacy cost annotation in the privacy function type to determine the privacy cost of the query and reduce the privacy budget accordingly.

\paragraph{Expressive Power of \duetsgx Queries.}
We have shown just a few simple examples of valid \duetsgx queries here, but the \duet language is designed to be flexible and support complex programs beyond simple analytics. For example, \duet supports looping constructs with differential privacy, and can use advanced composition to determine privacy cost for these programs; \duet also supports recent variants of differential privacy like R\'{e}nyi differential privacy~\cite{mironov2017renyi} and zero-concentrated differential privacy~\cite{bun2016concentrated}, and these can be used in \duetsgx queries to improve composition. In addition to programs that compute traditional analytics, we have also implemented \duetsgx queries for machine learning (e.g. noisy gradient descent algorithms~\cite{chaudhuri2011differentially, bassily2014private, abadi2016deep}).

% \section{Implementation \& Guarantees}
% \label{sec:implementation}

% \subsection{Implementation}

% \subsection{Ensuring Differential Privacy}

% - Ensuring differential privacy: use of Duet

% \subsection{Ensuring Confidentiality \& Integrity}

%  - Ensuring confidentiality of data: properties of the "protocol" (what gets encrypted, with what keys, etc)
%  - Ensuring confidentiality \& integrity of computation: use of SGX

\section{Limitations \& Future Work}
\label{sec:limitations}

Our proof-of-concept implementation of \duetsgx is a prototype, and should not be used in production to protect sensitive data. It has not been tested extensively, and may contain bugs that would allow attacks that reveal the encrypted data.

In addition, limitations in Graphene result in known issues with \duetsgx that could result in security vulnerabilities. In particular, Graphene's remote attestation features are still under construction; it is not yet possible to embed \duetsgx's public key in the attestation quote that is generated. Instead, our prototype implementation exposes the public key separately, by writing it to a file. This approach means that the untrusted HTTP server could \emph{modify} the public key before sending it to clients, which could enable an adversary who does not control the secure enclave to decrypt and read the submitted data. As Graphene's support for remote attestation improves, we plan to embed the public key in the quote itself, which is signed by the SGX processor's private key so that it cannot be modified by the untrusted components of the system.

Intel SGX has been the subject of a number of attacks~\cite{gotzfried2017cache, van2018foreshadow, weichbrodt2016asyncshock, schaik2020cacheout, sgaxe}. Many of these attacks allow a process outside the enclave to recover secrets stored inside the enclave memory, including private keys, and would enable an attacker to defeat the protections of \duetsgx. Most of the existing attacks have been addressed quickly by Intel, and can be mitigated by updating the processor's microcode, but this issue should be carefully considered before SGX is used in any system to protect truly sensitive data. We plan to adapt our implementation of \duetsgx to new implementations of secure enclaves as they appear---AMD's Secure Encrypted Virtualization, for example, will allow running unmodified programs---and we expect fewer realistic attacks as secure hardware matures.

Finally, our implementation of \duetsgx does not support the complete range of features of the underlying \duet system. In particular, the privacy budget for \duetsgx is stored in terms of values for $\epsilon$ and $\delta$; programs that leverage other variants of differential privacy (e.g. R\'{e}nyi differential privacy or zero-concentrated differential privacy) can be written as \duetsgx queries, but their privacy costs must be converted to $(\epsilon, \delta)$---\duetsgx is not currently capable of using these variants for sequential composition of queries. We plan to explore additional features for composition---including variants like these and also the ability to submit workloads of queries for improved accuracy.
 
\section{Related Work}

The use of SGX to provide security for outsourced computation has been studied extensively, and systems like Opaque~\cite{zheng2017opaque}, VC3~\cite{schuster2015vc3}, Haven~\cite{baumann2015shielding}, Ironclad Apps~\cite{hawblitzel2014ironclad}, and Ryoan~\cite{hunt2018ryoan} achieve this goal in various ways. These approaches typically assume the analyst to be trusted, and do not integrate differential privacy.

Ongoing research in the local model of differential privacy has produced a significant number of systems, including RAPPOR~\cite{erlingsson2014rappor}, Prochlo~\cite{bittau2017prochlo}, Apple's system~\cite{apple}, and many other works in various kinds of analytics~\cite{wang2017locally, qin2016heavy} and machine learning~\cite{duchi2013local}. Recent work has explored combining differential privacy with cryptographic techniques for secure computation: Honeycrisp~\cite{roth2019honeycrisp} uses a secure aggregation protocol to eliminate the trusted data curator for various analytics, and DJoin~\cite{narayan2012djoin}, Shrinkwrap~\cite{bater2018shrinkwrap}, and Crypt$\epsilon$~\cite{roy2020crypt} run a variety differentially private analytics queries using multiparty computation at a smaller scale. These systems do not require special hardware, but they introduce significant performance overhead and provide a less expressive query language than \duetsgx.

% - SGX + analytics (VC3, Opaque, etc)
% - SGX containers (Graphene, SCONE, etc)
% - Local DP systems (RAPPOR, Apple, etc)
% - Crypto approaches (Dwork, DJoin, Honeycrisp, Cryptepsilon, Xi He's recent work)

\section{Conclusion}

We have presented \duetsgx, a platform for collecting and processing sensitive data that uses secure hardware to guarantee confidentiality for the data and ensure differential privacy for all outputs. \duetsgx collects and stores encrypted data, and allows its decryption only within a secure enclave that executes \emph{only} differentially private queries. Queries are written using the \duet language and verified to be differentially private using the \duet typechecker. Our proof-of-concept implementation demonstrates that secure hardware can be used to build systems that provide the benefits of both the local and central models of differential privacy, without requiring a trusted data curator.

\section*{Acknowledgments}

We would like to thank the TPDP20 reviewers for their helpful suggestions for improvements.
This work was supported by
% David's NSF
NSF via award CCF-1901278, by 
% HECTOR
ODNI/IARPA via award 2019-1902070008,
and by DARPA \& SPAWAR under contract N66001-15-C-4066. The U.S. Government is authorized to reproduce and distribute reprints for Governmental purposes not withstanding any copyright notation thereon. The views, opinions, and/or findings expressed are those of the author(s) and should not be interpreted as representing the official views or policies of the Department of Defense or the U.S. Government.

\bibliographystyle{plain}
\bibliography{refs}

\end{document}